\documentclass[twocolumn]{article}        % twocolumn
\usepackage{graphicx}
\usepackage{epstopdf}
\usepackage{array}
\usepackage{url}
\usepackage{calc}
\usepackage{eurosym}
\usepackage{gensymb}
\usepackage{enumitem}  
\usepackage{amsmath}
\usepackage{amssymb}
\usepackage{mathtools}
\usepackage{multicol}
\usepackage{float}
\usepackage{subfig}
\usepackage{authblk}

\usepackage{geometry}
\title{Cost Optimal Design of Zero Emission Neighborhoods' (ZENs) Energy System: Model Presentation and Case Study on Evenstad}
\date{}

\begin{document}

\author[*]{Dimitri Pinel}
\author[*]{Magnus Korp{\r a}s}
\author[**]{Karen B. Lindberg}
\affil[*]{Dept. of Electric Power Engineering, NTNU, Trondheim, Norway}
\affil[**]{SINTEF Byggforsk, Oslo, Norway}

\maketitle

\begin{abstract}
Zero Emission Neighborhoods (ZEN) is a concept studied in Norway to reduce the $CO_2$ emission of neighborhoods.
One question coming along this concept is how to design the energy system of such neighborhoods to fit the ZEN definition\cite{zendef}. From this definition we extract the $CO_2$ balance, requiring an annual net zero emission of $CO_2$ in the lifetime of the neighborhood.
This paper proposes a MILP model for obtaining cost optimal design of ZEN's energy system and demonstrates it on a case study. Different technologies are included as investment options and, notably PV as an on-site electricity production mean. Wind turbines are not included in this study because inappropriate in the context of most cities.  
The results for the case study highlight the importance of PV investment in reaching the ZEN requirements. For example, around 850 kW of solar is needed for our test cases of $10 000 m^2$ of floor area, for an annual energy demand of around $700 MWh$ of electricity and $620 MWh$ of heat. The investments in other technologies are small in comparison.
\end{abstract}

\section{Introduction}
\label{intro}

A ZEN is a neighborhood that has a net zero emission of CO2 over its lifetime. Many aspects are embedded in the idea of ZEN. Energy efficiency, materials, users behaviour, energy system integration are all aspects that need to be accounted for in this concept. In addition, different part of the life cycle can be included but in this paper we only consider the operation phase and no embedded emission.

Two types of action exist to make a neighborhoods more sustainable. One is to act on the demand, via better insulation, user behavior or other efficiency measures. The other is to act on the supply and have a local energy system minimizing the $CO_2$ emissions.
There is consequently a need for a way of designing the energy system of such neighborhoods. The questions to be answered are, which technologies are needed to satisfy the demand of heat and electricity of a neighborhood, and how much of it should be installed so that it is as inexpensive as possible. The problem is then to minimize the cost of investment and operation in the energy system of a neighborhood so that it fulfills the ZEN criteria.
This paper presents an optimization model to solve such problems with a focus on operations research methodology.

\section{State of the Art and Contribution}
\label{sec:1}

The ZEN concept is specific to this particular project, however similar topics have been studied in different settings either at the neighborhood level, the city level or the building level, for example during the research center on Zero Emission Building.
In this context, K B Lindberg studied the investment in Zero Carbon Buildings \cite{ZCB_lindberg_14} and Zero Energy Buildings \cite{lindberg_methodology_2016} which are variations around the concept of ZEB. In both papers an optimization based approach is used to study the impact of different constraints on the resulting design. The second one (\cite{lindberg_methodology_2016}) in particular uses binary variables to have a more realistic representation of the operation part (part load limitation and import/export). 
In \cite{gabrielli_optimal_2018}, Gabrielli et al. tackle the problem of investment and operation of a neighborhood system and show an approach allowing to model the system complexity while keeping a low number of binary variables. It also constrains the total $CO_2$ emissions. It uses design days and proposes two methods for allowing to model seasonal storages while keeping the model complexity and reducing the run time.
In \cite{hawkes_modelling_2009}, Hawkes and Leach look at the design and unit commitment of generators and storage in a microgrid context using 12 representative days per season in a linear program. It is particular in that it defines how much the microgrid would be required to operate islanded from the main grid and include this in the optimization. It also discusses the problematic of market models within microgrids. 
In \cite{weber_optimisation_2011}, Weber and Shah present a mixed integer linear programming tool to invest and operate a district with a focus on cost, carbon emission and resilience of supply. A specificity of this tool is that it also designs the layout of the heat distribution network taking into account the needs of the buildings and the layout of each areas. It uses the example of a town in the United Kingdom for its case study.
In \cite{mehleri_mathematical_2012}, Mehleri et al. study the optimal design of distributed energy generation in the case of small neighborhoods and test the proposed solution on a Greek case. Emphasis is put on the different layouts of the decentralized heating network.
In \cite{schwarz_two-stage_2018}, Schwarz et al. present a model to optimize the investment and the energy system of a residential quarter, using a two stage stochastic MILP. It emphasizes on how it tackles the stochasticity of the problem in the different stages, from raw data to the input of the optimization, and on the computational performances and scalability of the proposed method.

In this paper, the focus is put on getting a fast yet precise solution that can take long term trends, such as cost reduction of technologies or climate. To this end, the proposed model uses a full year representation, ensuring a correct representation of seasonal storage of heat and electricity, and allows to divide the lifetime of the neighborhood into several periods, each represented by one year. It is also different by using the Zero Emission framework on a neighborhood level as a guide for the emission reduction constraint. This adds an integral constraints coupling each timestep and increasing the complexity of the problem. The use of binary variables is limited to the minimum.

\section{ZENIT Model Description}
\label{sec:2}

ZENIT stands for Zero Emission Neighborhoods Investment Tool. It is a linear optimization program written in Python and using Gurobi as a solver. It minimizes the cost of investing and operating the energy system of a ZEN using periods, with a representative year in each period. Different technologies are available, both for heat and for electricity. It is most suited for greenfield investment planning but can also take into account an existing energy system. The objective function is presented below:
\begin{equation}
    \begin{split}
        \MoveEqLeft
        \sum_{i} C_i^{disc}\cdot x_i + b_{hg} \cdot C_{hg} + \frac{1}{\varepsilon^{tot}_{r,D}} \sum_{i} C_i^{maint} \cdot x_i\\+ \sum_{p}  \varepsilon_{r,p}&\bigg(\sum_{t}\Big(\sum_{f}f_{f,t,p}\cdot P_{f,p}^{fuel} + (P_{t,p}^{spot}+P^{grid}\\+P^{ret})&\cdot (y_{t,p}^{imp}+\sum_{est}y_{t,p,est}^{gb\_imp})-P_{t,p}^{spot}\cdot y_{t,p}^{exp}\Big)\bigg)
    \end{split}
\end{equation}

The objective is to minimize the cost of investing in the energy system as well as its operation cost.

The operation phase can be separated in different periods during the lifetime of the neighborhood, and one year with hourly time-steps is used for each period. In addition to technologies producing heat or electricity, there is also the possibility to invest in a heating grid represented by the binary $b_{hg}$ that also gives access to another set of technologies that would be inappropriate at the building level.
In the equation above, the $\mathcal{E}$ represent discount factors either global for the whole study (\ref{disc_gl}) or for each period (\ref{disc_p}). They are calculated in the following way:
\begin{multicols}{2}\noindent
    \begin{equation}\label{disc_p}
        \varepsilon^{tot}_{r,D}= \frac{r}{1-(1+r)^{-D}}
    \end{equation}\noindent
    \begin{equation} \label{disc_gl}
        \varepsilon_{r,p}=\frac{(1+r)^{-p\cdot YR}}{\frac{r}{1-(1+r)^{-YR}}} 
    \end{equation}
\end{multicols}

The calculation assumes that reinvestment in this technology is made for the whole lifetime of the neighborhood, and is discounted to year 0. The salvage value is also accounted for. The formula used is :
\begin{equation}
\begin{split}
\MoveEqLeft
    C^{disc}_i = \bigg(\sum_{n=0}^{N_i-1} C^{inv}_i\cdot (1+r)^{(-n\cdot L_i)}\bigg) \\&- \frac{N_i\cdot L_i- D}{L_i}\cdot C^{inv}_i\cdot (1+r)^{-D}    
\end{split}
\end{equation}
\begin{equation}
    with:    N_i=\Bigg\lceil \frac{D}{L_i} \Bigg\rceil
\end{equation}

In the objective function, $y_{t,p}^{exp}$ represent the total export from the neighborhood. It is simply the sum of all exports from the neighborhood:
$\forall t,p$
\begin{equation}
    y_{t,p}^{exp}= \sum_{g}y_{t,p,g}^{exp}+\sum_{est}(y_{t,p,est}^{gb\_exp}+y_{t,p,est}^{pb\_exp})\cdot \eta_{est}
\end{equation}

The most important constraint, and what makes the specificity of the "Zero Emission" concept, is the $CO_2$ balance constraint. It is a net zero emission constraint of $CO_2$ over a year. This constraint is expressed below, $\forall p$:
\begin{equation}
    \begin{split}
    \MoveEqLeft
    \sum_{t}((y_{t,p}^{imp}+\sum_{est}y_{t,p,est}^{gb\_imp}) \cdot \varphi^{CO_2}_{e}) \\&+ \sum_{t}\sum_{f} (\varphi^{CO_2}_{f} \cdot f_{f,t,p}) \leq \sum_{t}(\sum_{est}(y_{t,p,est}^{gb\_exp}\\&+y_{t,p,est}^{pb\_exp})\cdot \eta_{est} +\sum_{g}y_{t,p,g}^{exp}) \cdot \varphi^{CO_2}_{e}
    \end{split}
\end{equation}

In the ZEN framework, this constraint is set as an an annual constraint. It can however also be used for shorter periods of time.

\begin{figure}
    \centering
    \includegraphics[width=0.48\textwidth]{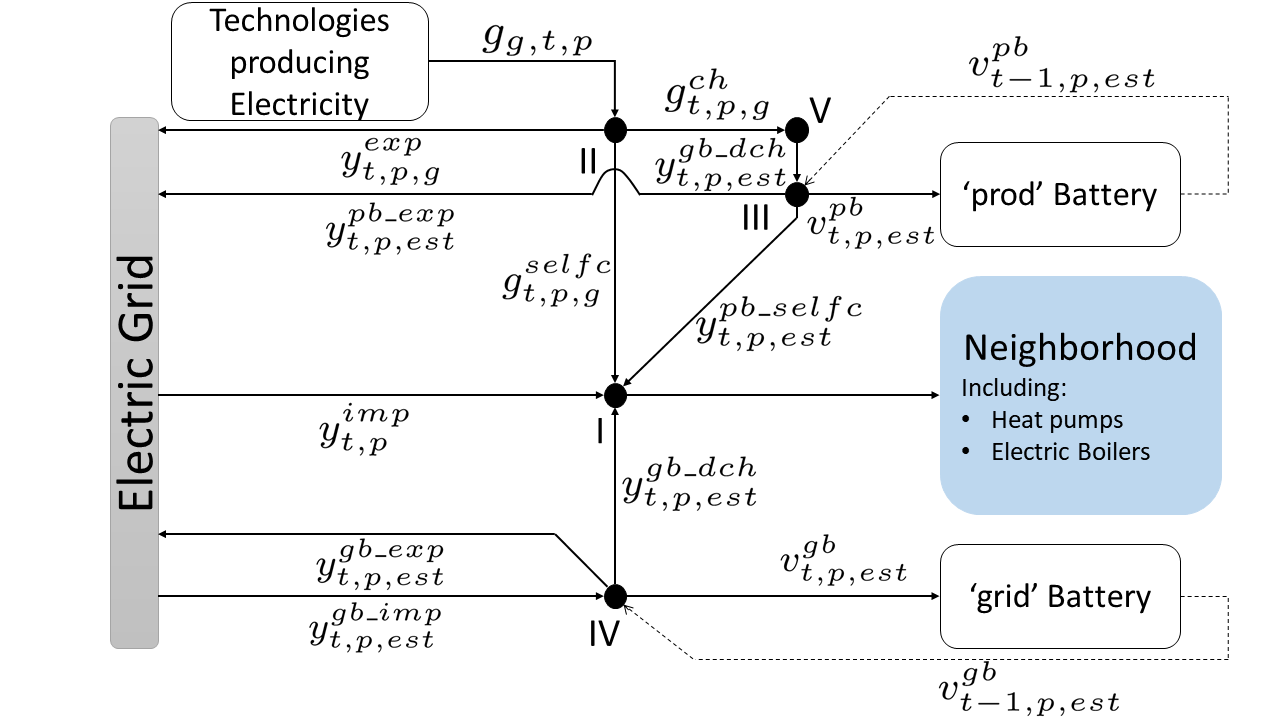}
    \caption{Graphical representation of the electricity balance in the optimization}
    \label{fig:elbal}
\end{figure}

Other necessary constraints are the different electricity and heat balances which guarantee that the different loads are served at all times. The electricity balance is represented graphically in Figure \ref{fig:elbal}. The corresponding equations are also written below. The electricity balance is particular because, we want to keep track of the origin of the electricity sent to the battery. It is managed by representing each battery as a combination of two other batteries: one is linked to the on-site production technologies, while the other is connected to the grid. It allows to keep track of the self-consumption and to differentiate between the origin of the energy for the $CO_2$ balance. 

Node I (\ref{eqI}) represents the main electric balance equ\-ation while II (\ref{eqII}) and V (\ref{eqV}) are only related to the on-site production of electricity. Node II (\ref{eqII}) describes that the electricity produced on-site is either sold to the grid, used directly or stored, while node V (\ref{eqV}) states that at a given time step what is stored in the batteries is equal to what is in excess from the on-site production.

Electricity balance I: $\forall t,p$
\begin{equation} \label{eqI}
    \begin{split}
    \MoveEqLeft
    y^{imp}_{t,p}+\sum_{est}(y^{gb\_dch}_{t,p,est}+y^{pb\_selfc}_{t,p,est}) \cdot \eta_{est} + \sum_{g} g_{g,t,p}^{selfc} \\&= \sum_e d_{e,t,p} + \sum_b \sum_{hp} d_{hp,t,p,b} + \sum_b E_{b,t,p} \cdot A_b
    \end{split}
\end{equation}

Electricity balance II: $\forall t,p,g$
\begin{equation}\label{eqII}
    g_{g,t,p}=y_{t,p,g}^{exp}+g_{g,t,p}^{selfc}+g_{t,p,g}^{ch}
\end{equation}

Electricity balance V: $\forall t,p$
\begin{equation}\label{eqV}
    \sum_{g}g_{t,p,g}^{ch}=\sum_{est}y^{pb\_ch}_{t,p,est}
\end{equation}

Heat also has its own balance, that guarantees that the demand of each building is met:
\begin{equation}
    \begin{split}
    \MoveEqLeft
    \sum_{\gamma \in \mathcal{Q} \smallsetminus \mathcal{HP}} q_{\gamma,t,p} + \sum_b \sum_{hp} q_{hp,t,p,b} \\&+ \sum_{hst}\eta_{hst} \cdot q^{dch}_{t,p,hst} = \sum_b H_{b,t,p} \cdot A_b +q^{ch}_{t,p}
    \end{split}
\end{equation}

The batteries are represented, as mentioned earlier and as seen on Figure \ref{fig:elbal}, as two entities: one on the on-site production side and the other on the grid side. This means that we have two "virtual" batteries with their own set of constraints as well as constraints linking the two.

The first constraint is a "reservoir" type of constraint and it represents the energy stored in the battery at each time-step: $\forall t\in \mathcal{T}^*,p,est$
\begin{equation}
    v^{pb}_{t,p,est}=v^{pb}_{t-1,p,est}+\eta_{est}\cdot y^{pb\_ch}_{t-1,p,est}-y_{t-1,p,est}^{pb\_exp}-y^{pb\_selfc}_{t-1,p,est}
\end{equation}
\begin{equation}
    v^{gb}_{t,p,est}=v^{gb}_{t-1,p,est}+\eta_{est}\cdot y_{t-1,p,est}^{gb\_imp}-y_{t-1,p,est}^{gb\_exp}-y^{gb\_dch}_{t-1,p,est}
\end{equation}
Equations \ref{e1}, \ref{e3} and \ref{e4} link both batteries. They make sure the sum of the stored energy in the "virtual" batteries is less than the installed capacity, and making sure the rate of charge and discharge of the battery is not violated. $\forall t,p,est$
\begin{equation}\label{e1}
     v^{pb}_{t,p,est}+v^{gb}_{t,p,est} \leq v^{bat}_{t,p,est}
\end{equation}
\begin{equation}\label{e2}
     v^{bat}_{t,p,est} \leq x_{bat,est}
\end{equation}
\begin{equation}\label{e3}
     y^{pb\_ch}_{t,p,est}+y_{t,p,est}^{gb\_imp} \leq \dot{Y}_{max,est}^{bat}
\end{equation}
\begin{equation}\label{e4}
     y^{gb\_dch}_{t,p,est}+y_{t,p,est}^{gb\_exp} \leq \dot{Y}_{max,est}^{bat}
\end{equation}

The storage level at the beginning and the end of the periods should be equal.  $\forall p,est$
\begin{equation}\label{e5}
    v^{bat}_{start,p,est}=v^{bat}_{end,p,est}
\end{equation}

The heat storage technologies also have the same kind of equations as the batteries, for example: 
$\forall t\in \mathcal{T}^*,p,hst$
\begin{equation}
    v^{heatstor}_{t,p,hst}=v^{heatstor}_{t-1,p,hst}+\eta^{heatstor}_{hst}\cdot q_{t,p,hst}^{ch} -q_{t,p,hst}^{dch}
\end{equation}
Equations \ref{e1} to \ref{e5} also have equivalents for the heat storages. However the heat storages are not separated in two virtual entities since there are no exports of heat from the building.

The power exchanges with the grid are limited depending on the size of the connection: $\forall t,p$
\begin{equation}
    (y_{t,p}^{imp}+y_{t,p}^{exp}+\sum_{est}y_{t,p,est}^{grid\_imp,bat}) \leq GC
\end{equation}
In order to not add additional variables, the mutual exclusivity of import and export is not explicitly stated. It is still met however due to the price difference associated with importing and exporting electricity.

In addition to the above equations, different constraints are used to represent the different technologies included. The maximum investment possible is limited for each technology. $\forall i$:
\begin{equation}
    x_i \leq X_{i}^{max}
\end{equation}
The amount of heat or electricity produced is also limited by the installed capacity:
\begin{multicols}{2}\noindent
\begin{equation}
   \forall q,t,p: q_{q,t,p} \leq x_q
\end{equation}\noindent
\begin{equation}
   \forall g,t,p: g_{g,t,p} \leq x_g
\end{equation}
\end{multicols}

The amount of fuel used depends on the amount of energy provided and on the efficiency of the technology: respectively $\forall \gamma \in \mathcal{F} \cap \mathcal{Q},p,t$ and $\forall \gamma \in \mathcal{E} \cap \mathcal{Q},p,t$
\begin{multicols}{2}\noindent
\begin{equation}
    f_{\gamma,t,p}=\frac{q_{\gamma,t,p}}{\eta_\gamma}
\end{equation}\noindent
\begin{equation}
    d_{\gamma,t,p}=\frac{q_{\gamma,t,p}}{\eta_\gamma}
\end{equation}
\end{multicols}

For CHPs technologies, the Heat to Power ratio is used to set the production of electricity based on the production of heat. $\forall t,p$
\begin{equation}
    g_{CHP,t,p}= \frac{q_{CHP,t,p}}{\alpha_{CHP}}
\end{equation}

For the heat pumps, the electricity consumption is  based on the coefficient of performance (COP).

$\forall hp,b,t,p$
\begin{equation}
    d_{hp,b,t,p}=\frac{q_{hp,b,t,p}}{COP_{hp,b,t,p}}
\end{equation}

The heat pumps are treated differently from the other technologies because they are not aggregated for the whole neighborhood but are separated for each buil\-ding. This is because the COP depends on the temperature to supply, which is different in passive buildings and in older buildings and which is also different for domestic hot water (DHW) and for space heating (SH), and dependent on the temperature of the source. The source is either the ground or the ambient air depending on the type of heat pump. The COP is then calculated using a second order polynomial regression of manufacturers data\cite{lindberg_methodology_2016}.
The global COP is calculated as the weighted average of the COP for DHW and SH.

The solar technologies, solar thermal and PV, also have their own set of specific constraints.
$\forall t,p$:
\begin{equation}
    g^{PV}_{t,p}+g^{curt}_{t,p} = \eta^{PV}_{t,p} \cdot x_{PV} \cdot IRR_{t,p}
    \label{pv}
\end{equation}
\begin{equation}
    q^{ST}_{t,p}= x_{ST} \cdot \frac{IRR_{t,p}}{G_{stc}}
    \label{st}
\end{equation}

The hourly efficiency of the PV system is calculated based on \cite{hellman14}, and accounts for the outside temperature and the irradiance.
This irradiance on a tilted surface is derived from the irradiance on a horizontal plane that is most often available from measurements sites by using the geometrical properties of the system: azimuth and elevation of the sun and tilt angle and orientation of the panels.

The irradiance on the horizontal plane data comes from ground measurements from a station close to the studied neighborhood which can for example be obtained from Agrometeorology Norway\footnote{\url{lmt.nibio.no}}. The elevation and azimuth of the sun is retrieved from an online tool\footnote{Sun Earth Tools: \url{https://www.sunearthtools.com/dp/tools/pos_sun.php}}.
This calculation takes into account the tilt of the solar panel and its orientation. Several assumptions were necessary to use this formula. Indeed, the solar irradiance is made up of a direct and an indirect part and only the direct part of the irradiance is affected by the tilt and orientation. However there is no good source of irradiance data that provides a distinct measurement for direct and indirect parts in Norway as far as the authors know. Thus we make assumptions that allow us to use the complete irradiance in the formula. We assume that most of the irradiance is direct during the day and that most is indirect when the sun is below a certain elevation or certain azimuths. This assumption gives a good representation of the morning irradiances while still accounting for the tilt and orientation of the panel during the day. On the other hand, this representation overestimates the irradiance during cloudy days, when it is mostly indirect irradiance.

\section{Implementation}
\label{sec:3}

The model presented in the previous section has been implemented in the case of campus Evenstad, which is a pilot project in the ZEN research center\cite{pilotproj}. This implementation of the model and the parameters used are presented in this section. 
Campus Evenstad is a university college located in southern Norway and is made up of around 12 buildings for a total of about 10 000 $m^2$. Most of the buildings were built between 1960 and 1990 but others stand out. In particular two small buildings were built in the 19th century and the campus also features two recent buildings with passive standards. The campus was already a pilot project in the previous ZEB centre and one of those buildings was built as a Zero Emission Building. In addition, on the heating side a 100kW CHP plant (40kW electric) and a 350kW Bio Boiler both using wood chips were installed along with 100$m^2$ of solar collectors, 10 000L of storage tank, 11 600L of buffer tank and a heating grid. On the electric side, the same CHP is contributing to the on-site generation as well as a 60kW photovoltaic system. A battery system is already planned to be built accounting for between 200 and 300 kWh. Based on this we assume in the study an existing capacity of 250kWh. We keep those technology in the energy system of the neighborhood for one part of the study.

The technologies included in the study are listed in table \ref{tab:tech} along with the appropriate parameters.

\begin{table}
    \caption{Technologies Used in the Evenstad Case and Their Main Parameters}
    \resizebox{0.47\textwidth}{!}{%
    \begin{tabular}{llll}
    \hline\noalign{\smallskip}
    Technology & Inv. Cost  & Life- & Efficiency \\
     & (\euro/kW)   & time &  (\%) \\
      &    & (Years)  &  \\
    \noalign{\smallskip}\hline\noalign{\smallskip}
    \textbf{Building}&  &  &     \\
    \noalign{\smallskip}\hline\noalign{\smallskip}
        PV & 1600  & 25 & 18 \\
        Solar Thermal & 700   & 25 & 70 \\
        Air source HP & 556 & 15 & $COP_t$ \\
        Ground source HP & 444  & 15 & $COP_t$ \\
        Biomass Boiler & 350  & 20 & 85 \\
        Electric Boiler & 750  & 30 & 100 \\
        Gas Boiler & 120 & 25 & 95 \\
    \noalign{\smallskip}\hline\noalign{\smallskip}
        \textbf{Neighborhood}   &  &  &     \\
    \noalign{\smallskip}\hline\noalign{\smallskip}
    Gas CHP & 739  & 25 & $45_{th};35_{el}$ \\
    Biomass CHP & 3300  & 25 & $40_{th};25_{el}$ \\
    Heat Pump & 660  & 25 & $COP_t$  \\
    Electric Boiler & 150  & 20 & 100  \\
    Gas Boiler & 60 & 25 & 95  \\
    \noalign{\smallskip}\hline
    \end{tabular}}
    \label{tab:tech}
\end{table}

Two main sources for the parameters and cost of the technologies are used as references for the study. Most of the technologies' data is based on a report made by the Danish TSO energinet and the Danish Energy Agency\cite{energinet} on technology data for energy plants.
The other source includes the technology data sheets made by IEA ETSAP\cite{etsap} and is used in particular for the gas and the biomass CHP.
The cost of PV is based on a report from IRENA\cite{irena17}.
The two efficiencies reported for the CHP plants correspond to the thermal and electrical efficiency, noted by a subscript ($_{th}$ for thermal and $_{el}$ for electrical).

\begin{table}
    \caption{Storage Technologies Used in the Evenstad Case and Their Main Parameters}
    \resizebox{0.47\textwidth}{!}{%
    \begin{tabular}{llllll}
    \hline\noalign{\smallskip}
    Technology & Inv. Cost  & Lifetime & Efficiency \\
     & (\euro/kWh)  & (Years) &  (\%) \\
      &    &  &  \\
    \noalign{\smallskip}\hline\noalign{\smallskip}
        Battery & 350 & 15 & 94     \\
        Heat Storage & 75  & 20 & 95   \\
    \noalign{\smallskip}\hline
    \end{tabular}}
    \label{tab:stortech}
\end{table}

The heat storage values are based on a data sheet by ETSAP\cite{etsap} while the values used for the batteries are based on a report from IRENA\cite{irenabat17}.

\begin{table}
    \caption{Fuel Cost and $CO_2$ Factors}
    \resizebox{0.47\textwidth}{!}{%
    \begin{tabular}{lll}
    \hline\noalign{\smallskip}
    Fuel & Cost (\euro/kWh) & $CO_2$ Factor ($gCO_2$/kWh) \\
    \noalign{\smallskip}\hline\noalign{\smallskip}
    Gas & 0.055 & 277\\
    Biomass & 0.041 & 7 \\
    Electricity & $P_{t,p}^{spot}$ & 17 \\
    \noalign{\smallskip}\hline
    \end{tabular}}
    \label{tab:fuel}
\end{table}

The values in table \ref{tab:fuel} come from different sources. The cost of biomass comes from EA Energy Analyses \cite{eaea13}, the cost of gas is based on the cost of gas for non household consumers in Sweden\footnote{\url{http://ec.europa.eu/eurostat/statistics-explained/index.php?title=File:Gas_prices_for_non-household_consumers,_second_half_2017_(EUR_per_kWh).png}}. For the technologies in table \ref{tab:tech}, the O\&M costs, expressed as a percentage of the investment costs, are respectively:1, 1.3, 1, 1.3, 2, 0.8, 2.3, 4, 5.5, 1, 1 and 5. For the storage technologies in table \ref{tab:stortech}, the operating cost is 0.
The $CO_2$ factors of gas and electricity for Norway are based on a report from Adapt Consulting\cite{adapt13} and the $CO_2$ factor for biomass is based on \cite{dokka13}.

The electricity prices for Norway are based on the spot prices for the Oslo region in 2017 from Nordpool\footnote{\url{https://www.nordpoolgroup.com/Market-data1/Dayahead/Area-Prices/ALL1/Hourly/?view=chart}}. On top of the spot prices, a small retailer fee and the grid charges are added\footnote{\url{https://www.nve.no/energy-market-and-regulation/network-regulation/network-tariffs/statistics-on-distribution-network-tariffs/}}.
The prices are rather constant with a fair amount of peaks in the winter and some dips in the summer.

The irradiance on the horizontal plane and temperatures are obtained and used in the calculations as described in the previous section. The ground station used to retrieve data is F{\r a}vang, situated 50 km to the west of Evenstad.
The electric and heat load profiles for the campus are derived from \cite{lindberg_impact_2017}. The load profiles are based on the result of the statistical approach used in these papers and the ground floor area of each type of building on the campus.
In addition, the domestic hot water (DHW) and Space Heating (SH) are derived from the heat load based on profiles from a passive building in Finland where both are known \cite{pal_energy_2016}.

\section{Results}
\label{sec:4}

\begin{figure}
    \centering
    \includegraphics[width=0.47\textwidth,trim=4 4 4 4,clip]{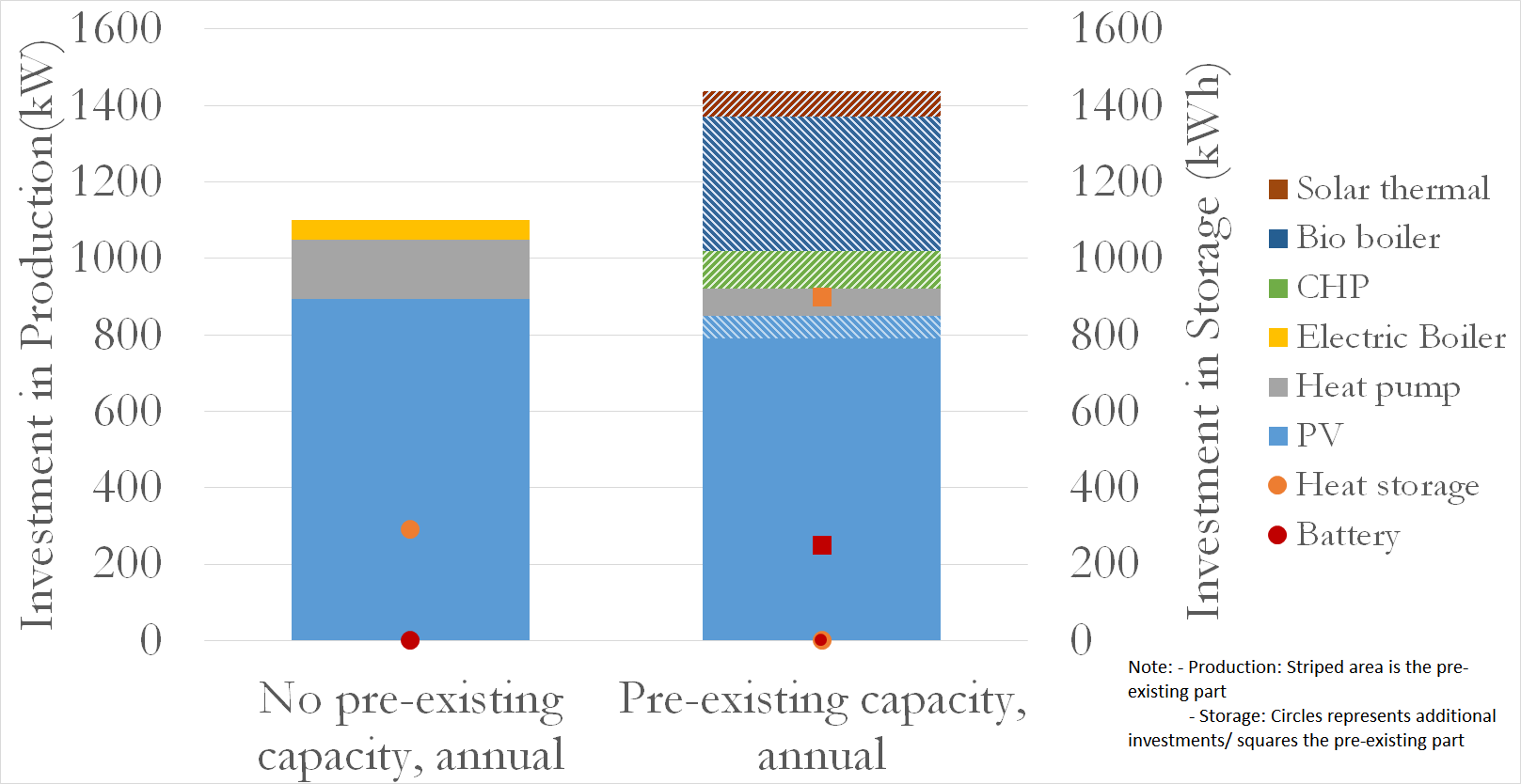}
    \caption{Resulting Energy System}
    \label{fig:y_result}
\end{figure}

The optimization was run several times with different conditions. It was run with a yearly $CO_2$ balance with and without including the energy system that already exists at Evenstad. When the pre-existing energy system is included, the amount of heat storage, PV, solar thermal and biomass heating (CHP and boilers) represents the minimum investment in the technology for the optimization. The energy systems resulting from those optimizations are presented on Figure \ref{fig:y_result}.

Both cases are interesting. Indeed the case with the pre-existing technologies included in the optimization allows to know in which technology to invest to move towards being a ZEN for the campus Evenstad while the case that does not include the pre-existing technologies allows to see how it would look like if it was built today from the ground up using the optimization model presented here and the given ZEN restrictions.

A first observation from Figure \ref{fig:y_result} is that the technologies already installed (Heat storage ST, Biomass Boiler BB, CHP, Battery) are not invested in more than what is already there, except for PV which gets a lot of additional investments to meet the ZEN criteria. In addition to the large investment in PV the only additional investment for Evenstad appears to be a heat pump.
In the case without any pre-installed technologies the system is quite different. There is still a need for investment in PV, though it is slightly lower and the optimization does not chose to invest in a battery. On the heating part the chosen design uses heat pumps and electric boiler in addition to a heat storage smaller than already installed in Evenstad.

Other system designs would have been possible in order to satisfy this constraint, based on biomass CHP for instance, but this appears to be be more expensive.

\begin{figure}
    \centering
    \subfloat[Summer]{
    \includegraphics[width=0.224\textwidth]{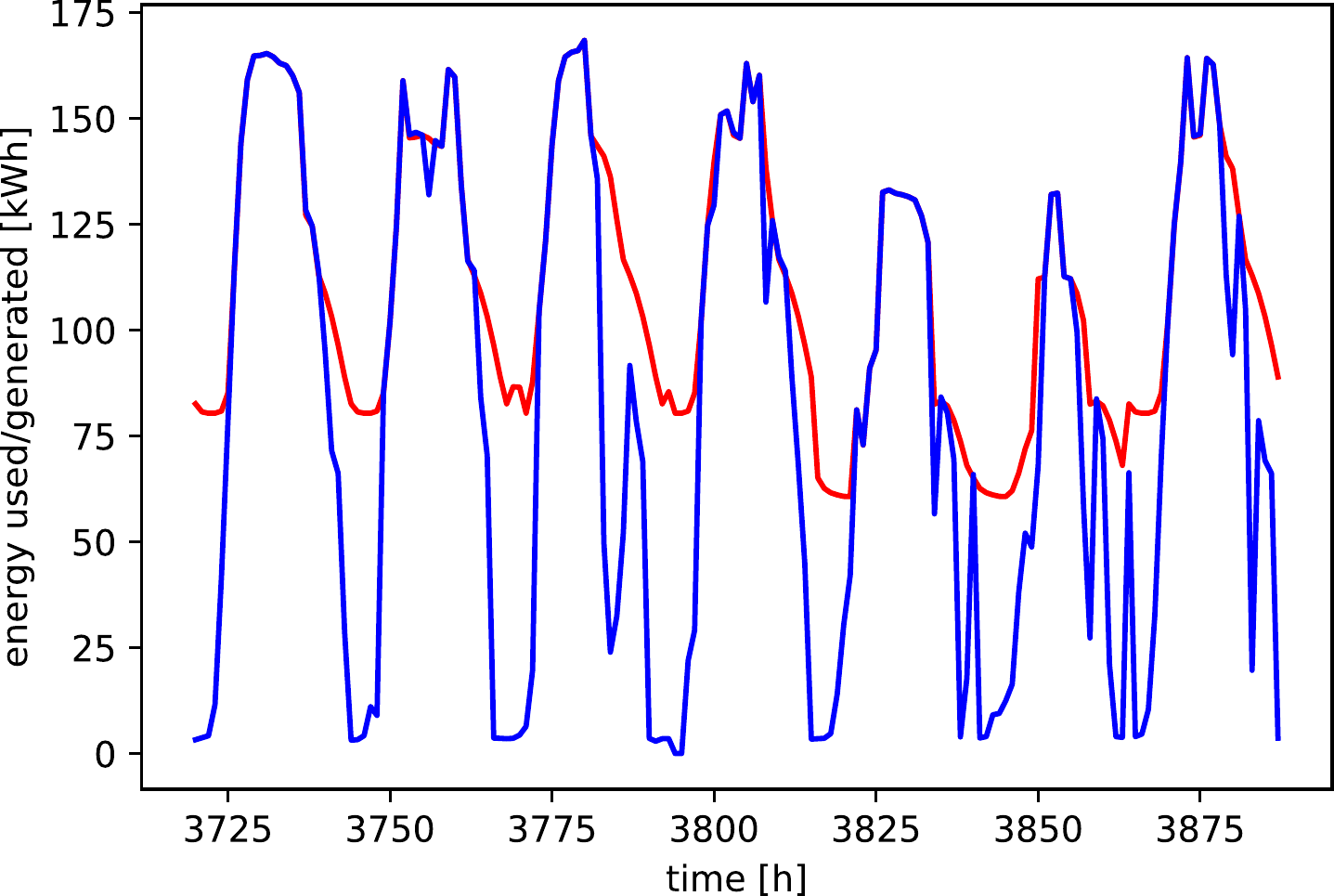}
    }
    \subfloat[Winter]{
    \includegraphics[width=0.224\textwidth]{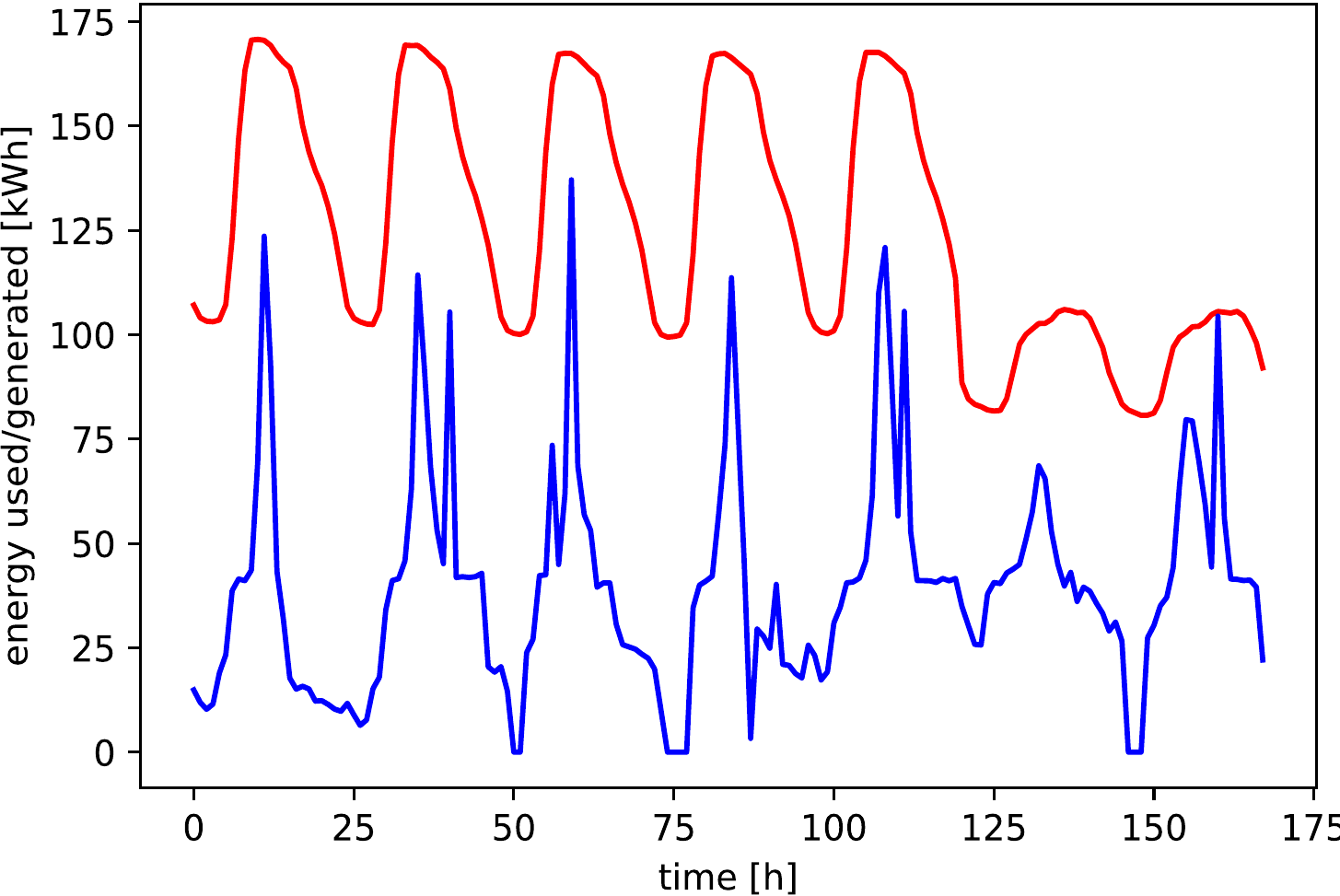}
    }
    \caption{Self Consumed Electricity (blue) and Total Consumption (red) of Electricity in the ZEN}
    \label{fig:selfc}
\end{figure}

\begin{figure}
    \centering
    \subfloat[Summer]{
    \includegraphics[width=0.224\textwidth]{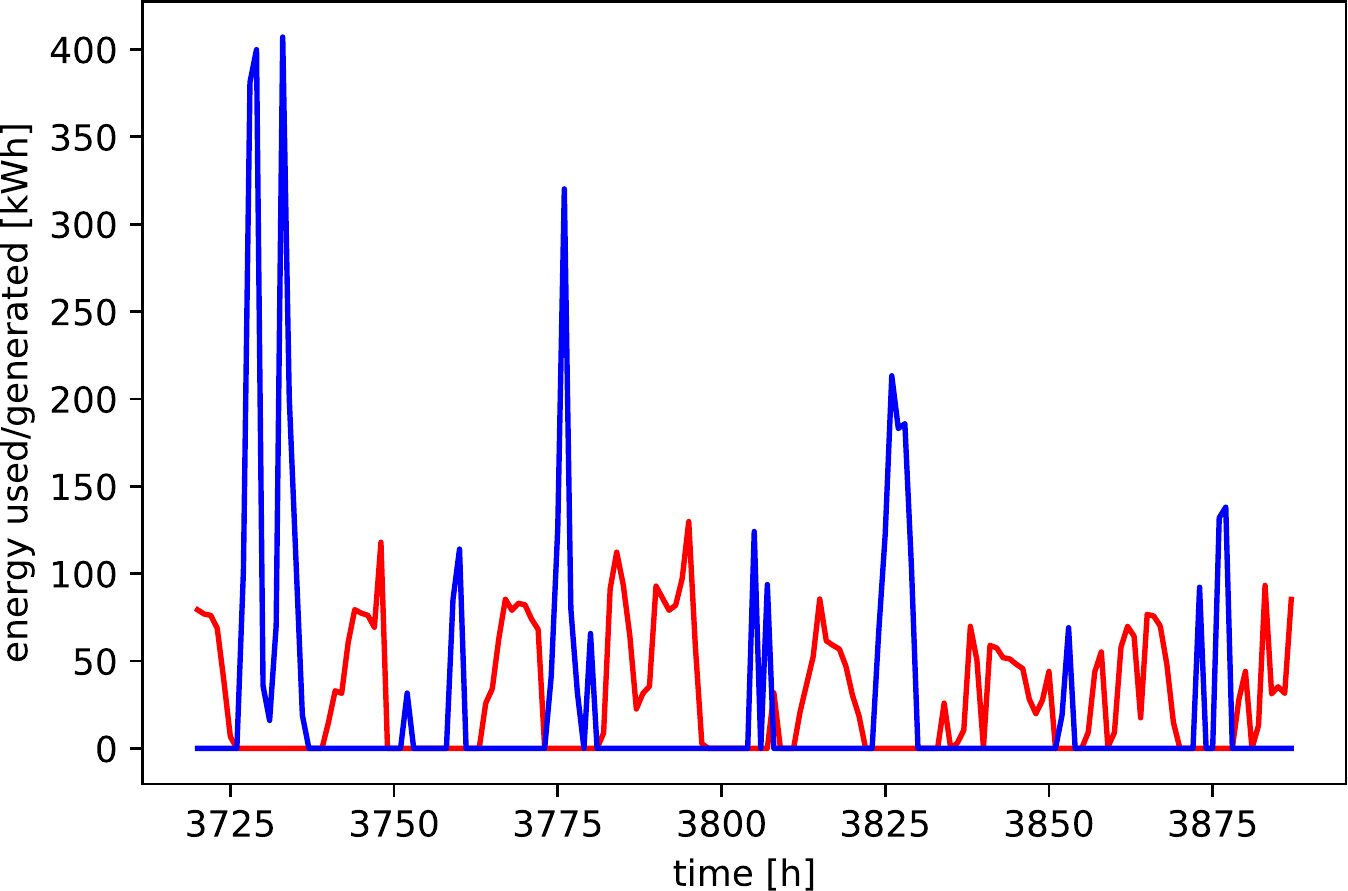}
    }
    \subfloat[Winter]{
    \includegraphics[width=0.224\textwidth]{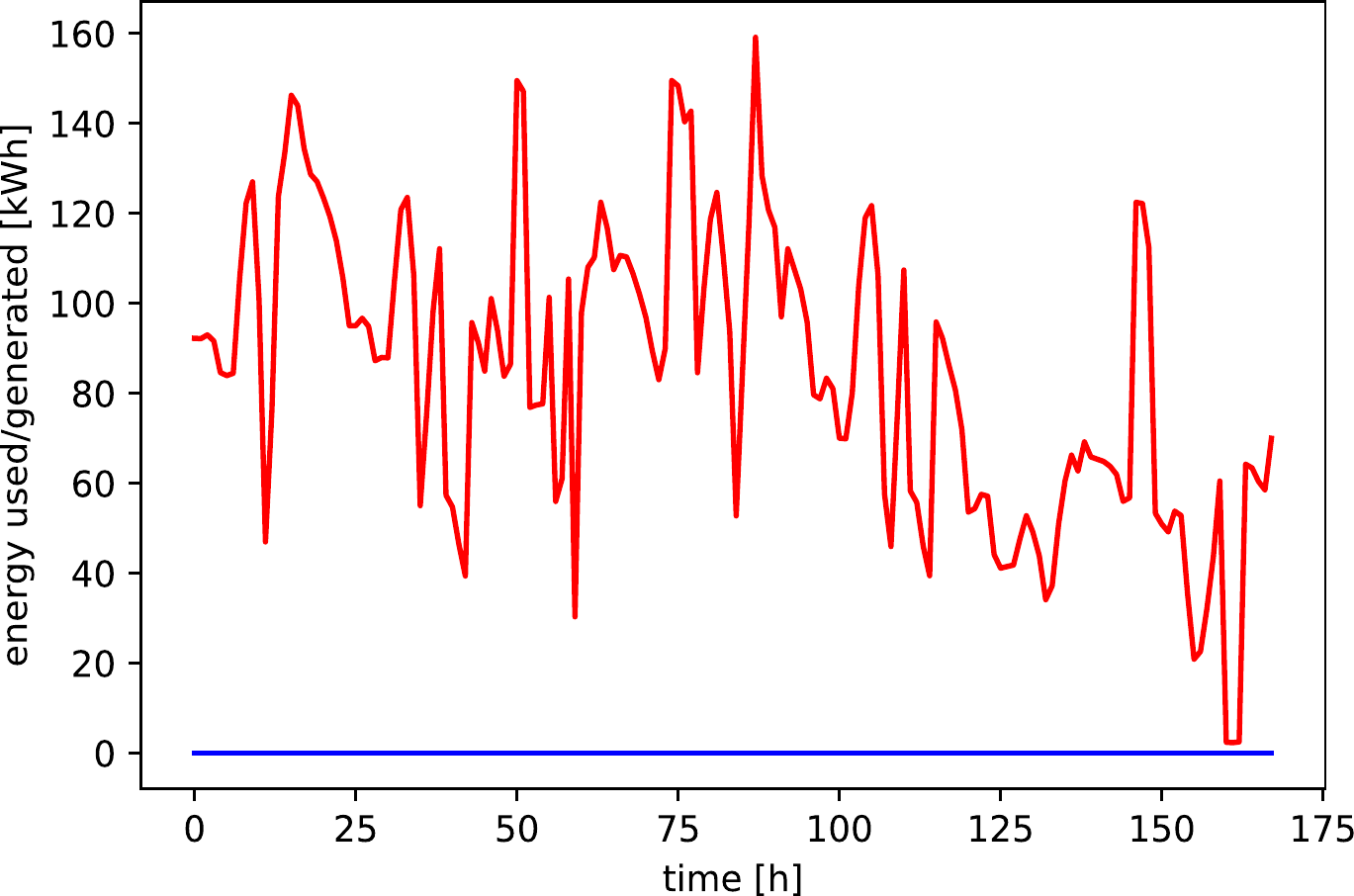}
    }
    \caption{Import (red) and Export (blue) of Electricity from the ZEN }
    \label{fig:impexp}
\end{figure}

On Figure \ref{fig:selfc}, the self consumption and the total demand of electricity is presented while on Figure \ref{fig:impexp} it is the imports (red) and exports (blue) of electricity that are presented. Both figures show a week for the case of the yearly balance and including pre-existing technologies. In the summer the neighborhood produces electricity in excess and needs to send it to the grid. The battery, that is part of the pre-existing technologies, is used but is not large enough to allow for relying on self produced electricity during the night. It is also not large enough to limit the amount of electricity sent to the grid. Figure \ref{fig:impexp} (a) illustrates this: the exports during the days have highs peaks that represent around four times the night imports in terms of peak power. This has implications on the sizing of the connection to the grid and is especially important in the context of the introduction of new tariffs based on peak power in Norway.
In the winter, some of the electricity is still self consumed due to the CHP that is part of the pre-existing technologies. This self consumption stays limited and no electricity is exported.

Ultimately, all resulting designs require huge investment in PV to attain the status of ZEN. In those systems, which rely heavily on electricity, heat pumps and electric boilers appear to be the preferred heating solution. 

\section{Limitations}

This study has several limitations, on the methodology and on the case study. For the case study, assumptions were necessary due to the lack of data, in particular for the loads or the insolation (diffuse and direct).
For the methodology, the will to limit the use of binary variables, left out constraints such as part load limitations which would be needed for some technologies. In addition, using an hourly resolution leads to an underestimation of the storage and , potentially, heating technologies size. There is a trade off between the solving time and the precision of the results and the resolution needs to be chosen accordingly.
Despite those limitations it provides insights in the design methodology that can be used to design the energy system of a ZEN.
The choice of $CO_2$ factors for electricity is also greatly impacting theresults and this should be studied in more detail in future work.

\section{Conclusion}

This paper presented in detail the ZENIT model for investment in Zero Emission Neighborhoods as well as its implementation and the results on a realistic case study of campus Evenstad in Norway, with a focus on operation research methodology.
The model is formulated as a MILP, using as few binaries as possible. The Zero Emission constraint complexify the problematic of designing the energy system of a neighborhood and the long term trends can be accounted for by defining periods.
For Evenstad, the results suggest that additional investments, mainly in PV, are necessary in order to attain the status of ZEN. Investments happen at both level but mainly at the building level.
When the technology already installed at Evenstad are not included, those technologies are not chosen by the investment (except for heat storage). The optimal choice in order to become zero emission for Evenstad in the current ZEN framework thus appears to be a massive investment in PV and a heating system fueled by electricity.
Further work includes disaggregating the heat part of the model and a more detailed operation part in the optimization. 
There are are key takeaways for policy makers in this study. The methodology presented in this paper can be used to assess policies such as design of grid tariffs or incentives on specific technologies in particular cases with a goal of zero emission.
And Under the $CO_2$ factor assumption used in this study, huge investment in PV are made which would be problematic in case of a large scale application of the concept of ZEN. This suggests the need for incentives in alternative technologies such as CHPs.

\section{Acknowledgements}
This article has been written within the Research Centre on Zero Emission Neighbourhoods in Smart Cities (FME ZEN). The authors gratefully acknowledge the support from the ZEN partners and the Research Council of Norway.

\section*{Nomenclature}
\begin{footnotesize}

\begin{description}[leftmargin=!,labelwidth=0pt]
\item[\textbf{Indexes (Sets)}] \  \\

\begin{description}[leftmargin=!,labelwidth=\widthof{\bfseries $q_{max}$}]
\item[$t (\mathcal{T})$] Timestep in hour within year $\in [0,8759]$
\item[$b (\mathcal{B})$] Building type 
\item[$yr$] Year within period $\in [1,YR]$
\item[$p$] Period
\item[$i (\mathcal{I})$] Energy technologies, $\mathcal{I} = \mathcal{F} \cup \mathcal{E} \cup \mathcal{HP} \cup \mathcal{S} \cup \mathcal{QST} \cup \mathcal{EST}; \mathcal{I} = \mathcal{Q} \cup \mathcal{G}$: \begin{description}[leftmargin=!,labelwidth=\widthof{$estaaaaaa$}]
\item[$f (\mathcal{F})$] Technology consuming fuel (gas, biomass, ...)
\item[$e (\mathcal{E})$] Technology consuming electricity
\item[$hp (\mathcal{HP})$] Heat pumps technologies
\item[$s (\mathcal{S})$] Solar technologies $\in {ST,PV}$
\item[$qst (\mathcal{QST})$] Heat storage technologies
\item[$est (\mathcal{EST})$] Electricity storage technologies
\end{description}
\item[$q (\mathcal{Q})$] Technologies producing heat
\item[$g (\mathcal{G})$] Technologies producing electricity
\end{description}

\item[\textbf{Parameters}] \  \\
\begin{description}[leftmargin=!,labelwidth=\widthof{\bfseries $COP_{hp,b,t,p}$}]
\item[$C_i^{disc}$] Discounted investment cost, technology $i$ with re-investments and salvage value [\euro/kWh]
\item[$\varepsilon_{r,p}$] Discount factor, period p with discount rate r
\item[$D$] Duration of the study [yr] : $D=P*YR$ 
\item[$P$] Number of periods in the study [-]
\item[$C_i^{maint}$] Annual maintenance cost [\% of inv. cost]
\item[$P_{f,p}^{fuel}$] Price of fuel of technology g, period p [\euro/kWh]
\item[$P_{t,p}^{spot}$] Electricity spot price [\euro/kWh]
\item[$P^{grid}$] Electricity grid tariff, period p [\euro/kWh] 
\item[$P^{ret}$] Retailer tariff on electricity, period p [\euro/kWh]
\item[$\eta_{est}$] Charge/Discharge efficiency of battery $est$ [-]
\item[$\varphi^{CO_2}_{e}$] $CO_2$ factor of electricity [g/kWh]
\item[$\varphi^{CO_2}_{f}$] $CO_2$ factor of fuel f [g/kWh]
\item[$\alpha_{CHP}$] Heat to power ratio of the CHP  [-]
\item[$GC$] Size of the neighborhood grid connection [kW]
\item[$X_{i}^{max}$] Maximum possible installed capacity of technology i [kW]
\item[$E_{b,t,p}$] Electric specific load of building b in timestep t in period p [$kWh/m^2$]
\item[$A_b$] Aggregated area of building b in the neighborhood [$m^2$]
\item[$H_{b,t,p}$] Heat specific load of building b in timestep t in period p [$kWh/m^2$]
\item[$\eta_{i}$] Efficiency of technology i [-]
\item[$COP_{hp,b,t,p}$] Coefficient of performance of heat pump hp in building b in timestep t in period p [-]
\item[$\dot{Y}_{max}^{bat}$] Maximum charge/dis- rate of battery [kWh/h]
\item[$\dot{Q}_{max}^{heatstor}$] Maximum charge/discharge rate of heat storage [kWh/h]
\item[$\eta^{PV}_{t,p}$] Efficiency of the solar panel in timestep t in period p [-]
\item[$L_{i}$] Lifetime of technology i [yr]
\item[$C_{hg}$] Cost associated with a heating grid for the neighborhood [\euro]
\end{description}

\item[\textbf{Variables}] \  \\
\begin{description}[leftmargin=!,labelwidth=\widthof{\bfseries $y^{pb\_selfc}_{t,p,est}$}]
\item[$x_{i}$] Capacity of technology $i$: for $i \in \{f \cup e \cup h \cup s\}$   [kW]; for $i \in \{qst \cup est\}$   [kWh]
\item[$f_{f,t,p}$] Fuel consumed by technology f in hour t [kWh]
\item[$d_{e,t,p}$] Electricity consumed by technology e in timestep t [kWh]
\item[$d_{hp,b,t,p}$] Electricity consumed by the heat pumps hp, in building type b [kWh]
\item[$y_{t,p}^{imp}, y_{t,p}^{exp}$] Electricity imported/exported from the grid to the neighborhood at timestep t [kWh]
\item[$y_{t,p,g}^{exp}$] Electricity exported by the production technology g to the grid at timestep t [kWh]
\item[$g_{t,p,g}^{selfc}$] Electricity generated from the technology g self consumed in the neighborhood, timestep t [kWh]
\item[$g_{t,p,g}^{ch}$] Electricity generated from the  technology g into the 'prod' batteries at timestep t [kWh]
\item[$y_{t,p,est}^{\_\_\_}$] Electricity imp/exported by the battery $est$ at timestep t [kWh] (gb\_exp, gb\_imp or pb\_exp) 
\item[$g_{g,t,p}$] Electricity generated by technology g in timestep t of period p [kWh]
\item[$q_{q,t,p}$] Heat generated by technology q in timestep t of period p [kWh]
\item[$y^{gb\_dch}_{t,p,est}$] Electricity discharged from the 'grid' battery $est$ to the neighborhood at timestep t [kWh]
\item[$y^{pb\_ch}_{t,p,est}$] Electricity charged from the neighborhood to the 'prod' battery $est$ at timestep t [kWh]
\item[$y^{pb\_selfc}_{t,p,est}$] Electricity to the neighborhood from the 'prod' battery $est$, timestep t [kWh]
\item[$q^{ch}_{t,p}$] Heat "charged" from the neighborhood to the heat storage at timestep t [kWh]
\item[$q^{dch}_{t,p}$] Heat "discharged" from the neighborhood to the heat storage at timestep t [kWh]
\item[$q^{ch}_{t,p}, q^{dch}_{t,p}$] Heat "charged"/"discharged" from the neighborhood to the heat storage at timestep t [kWh]
\item[$v^{gb}_{t,p,est}, v^{pb}_{t,p,est}$] 'grid'/'prod' Battery $est$ level of charge at timestep t in period p [kWh]
\item[$v^{heatstor}_{t,p}$] Heat storage level at timestep t in period p [kWh]
\item[$g^{curt}_{t,p}$] Solar energy production curtailed [kWh]
\item[$b_{hg}$] Binary variable for investment in a heating grid

\end{description}
\end{description}
\end{footnotesize}

\bibliographystyle{unsrt}
\bibliography{biblio.bib}  
\end{document}